\documentclass[11pt]{article}

\usepackage{anysize}

\renewcommand{\S}{{\cal S}}
\newcommand{\K}{{\cal K}}

\newcommand{\T}{{\cal T}}
\newcommand{\E}{{\cal E}}
\newcommand{\C}{{\cal C}}
\newcommand{\V}{{\cal V}}
\newcommand{\CP}{{\cal C}^\perp}
\newcommand{\VP}{{\cal V}^\perp}

\title{\bf New encoding schemes for quantum authentication}

\author{Priscila Garc\'{\i}a-Fern\'andez${}^1$ \and Enrique
Fern\'andez-Mart\'{\i}nez${}^1$ \and Esther P\'erez${}^2$ \and
David J. Santos${}^2$}

\date{${}^1$Instituto de \'Optica, CSIC, Serrano 123, E-28006 Madrid, Spain\\
${}^2$ETSIT, Universidad de Vigo, Campus Universitario s/n, E-36200 Vigo,
Spain\\}

\begin{document}

\maketitle

\begin{abstract}
We study the potential of general quantum operations, Trace-Preserving
Completely-Positive Maps (TPCPs), as encoding and decoding mechanisms in
quantum authentication protocols. The study shows that these general
operations do not offer significant advantage over unitary encodings.
We also propose a practical authentication protocol based on the use of
two successive unitary encodings.
\end{abstract}

\section{Introduction}

The past two decades have seen an enormous increase in the development
and use of networked and distributed systems, providing increased
functionality to the user and a more efficient use of resources. To
benefit from such systems, users cooperate by exchanging information over
communication channels. In many occasions, this information requires to
be protected from unauthorised users. The whole discipline of Cryptography
\cite{MENEZES_1996} addresses this issue.

Security in communication channels has been greatly improved by
recent advances in the field of Quantum Information Processing
(QIP). The most notable example is Quantum Key Distribution
(QKD) \cite{BENNETT_1984,EKERT_1991}, for which the phase of
experimental demonstration inside \cite{BENNETT_1992} and outside
\cite{MULLER_1995,MULLER_1996} research laboratories has lead to
the development of commercial products nowadays. One aspect of QKD
that is sometimes underestimated is the dependence of its security
on the existence of an authenticated classical channel between the
users. Basically, some technique must be designed to allow one party
(the verifier) to gain assurances that the identity of another (the
claimant) is as declared, thereby preventing impersonation. The most
common technique is by the verifier checking the correctness of a
message which demonstrates that the claimant is in possession of a
secret associated by design with the genuine party. Such a technique is
called a Message Authentication Code (MAC), first proposed by Gilbert
and co-workers in \cite{GILBERT_1974}.  More formally, an authentication
code involves a set of source messages, $\S$, a set of authentication
tags, $\T$, a set $\K$ of secret keys, and a set of rules such that each
$k\in \K$ defines a mapping to obtain the tag from the source message:
$u_k: \S \rightarrow \T$. According to this, an authenticated message,
consisting of a source state and its tag, $(s,t)$, with $t=u_k(s)$,
can be verified only by the intended recipient, with whom a key has
been shared previously. There are many MACs with different degrees of
security. MACs intended to provide unconditional security were first
studied by Wegman and Carter \cite{WEGMAN_1981} combining one-time
pads with hash functions. This approach was later pursued by Brassard
\cite{BRASSARD_1983} trading unconditional security for short keys,
and further refined by Krawczyk \cite{KRAWCZYK_1994}.

The use of quantum resources to obtain more efficient classical-message
authentication schemes is still an open issue. In \cite{CURTY_2001}
the authors showed that quantum information techniques can be used
to authenticate a binary classical message with a key of length
shorter than the one required by classical schemes. However, it is
not clear yet whether that would be the case with longer messages.
Leung \cite{LEUNG_2002} has recognised the potential relevance of
authenticating quantum information in future quantum communication
systems, and Barnum and coworkers \cite{BARNUM_2002} have proposed
a secret-key quantum authentication protocol that uses stabiliser
purity testing codes.  Surprisingly, they showed that any protocol that
guarantees secure authenticity must encrypt the quantum plain-text almost
perfectly.  This fact contrasts with classical MACs, where encryption
of the plain-text is not necessary for unconditional security. More
recently, Gea-Banacloche \cite{GEA_BANACLOCHE_2002} has approached data
authentication from a steganographic perspective, making use of quantum
correcting code techniques.

In this paper we extend and further develop previous work
\cite{CURTY_2002,PEREZ_2003} on the authentication of quantum information
with unitary coding sets. Our purpose is two-fold. First, in Section
\ref{AUTEN_TPCP}, we investigate whether the use of reversible TPCP
maps in the encoding and decoding stages of our protocol improves its
robustness against several common attacks; we restrict ourselves to the
simple case of a single qubit since our goal is to compare our results
with those of \cite{CURTY_2002}. Second, in Section \ref{AUTEN_DOBLE} we
study the authentication of arbitrarily complex quantum messages. Now,
instead of using a TPCP map, we propose a sequence of two unitary
encodings. We show that proceeding in such a way improves the security
of the authentication protocol. Finally, in Section \ref{FINALE}, we
present our main conclusions.

\section{Authentication with general quantum operations}
\label{AUTEN_TPCP}

In the general quantum authentication setting Alice sends a quantum
message to Bob with the goal of making Bob confident about the
authenticity of such message. If we consider that both participants share
a quantum secret key (for example, a set of EPR pairs), and they have
access to an authenticated classical channel, then the solution is quite
simple: Alice can just use quantum teleportation \cite{BENNETT_1993}
to send the quantum plain-text. However, here we shall assume that no
classical authenticated channel is available, and that the previously
shared keys are standard classical bits (the nature of this assumption is 
practical, given the technological difficulties that still exist
to the the manipulation of qubits).  Our main goal is to extend the
standard MAC technique to the quantum realm.  For the sake of simplicity,
let us assume that the source quantum message is a qubit described by
the density operator $\rho_\S$ belonging to a two-dimensional source
message space $\S$.  Following the standard procedure, Alice appends
a tag (a public-known quantum state) to the message in such a way
that the recipient, Bob, may verify the tag and so convince himself
about the identity of the message originator.  The tag is also
given by a density operator $\rho_\T$, belonging to a tag space
$\T$. Therefore, the quantum tagged-message is described by the
operator $\rho_\E=\rho_\S\otimes\rho_\T$ that acts on the state space
$\E=\S\otimes\T$. To validate the message, Alice and Bob also openly
agree on a particular splitting of the tag space $\T$ into valid and
invalid tags.  The space $\T$ can then be seen as the direct sum of
two subspaces, $\T=\V\oplus\VP$, where $\V$ is the subspace of valid
tags and $\VP$ is the subspace of the invalid ones.  The splitting of
$\T$ leads to the direct sum $\E=\C\oplus\CP$, where $\C=\S\otimes \V$
(${\rm dim}(\C)=C$) and $\CP=\S\otimes \VP$ (${\rm dim}(\CP)=D$) are,
respectively, the subspaces of valid and invalid tagged messages.
A message originally prepared by Alice will thus belong to $\C$.

We shall assume that Alice and Bob share a minimum-size key of just one
bit, $k$. Depending on its value, Alice performs an encoding rule on
the tagged message: She will apply an $I$ operation in one case (Alice
does nothing to the tagged message), and a reversible (to guarantee
perfect deterministic decoding) TPCP map \cite{NIELSEN_1998,CAVES_1999}
in the other. It is worth noting that this is completely equivalent,
in security terms, to the use of two TPCP maps. The action of the TPCP
map on $\rho_\E$ can be written as

\begin{equation}
E(\rho_\E)=\sum_j d_j U_j \rho_\E U^\dagger_j,
\label{TPCP}
\end{equation}

\noindent where the $U_j$ are unitary operators, the $d_j$ are real and
positive, and the following restrictions apply:

\begin{eqnarray}
\sum_jd_j=1, \label{cond1}\\
P_i U^\dagger_k U_j P_i=\delta_{jk}, \label{cond2}
\end{eqnarray}

\noindent where $P_i$ is the projector onto $\C$. Condition (\ref{cond1})
together with the unitarity of the $U_j$ operators guarantees that the
operation will be trace-preserving. Condition (\ref{cond2}) assures
the reversibility of the TPCP map if the input state $\rho_\E$ belongs
to the subspace of valid messages $\C$.  After the encoding, Alice
will send to Bob, depending on the value of the key, either $\rho_\E$
or $E(\rho_\E)$. In both cases we will call the message sent by Alice
$\rho_\E(k)$. On the reception side, Bob will decode the message sent
by Alice according to the value of the the key he also holds: If Alice
sent $\rho_\E$, Bob will not do anything to the message. Otherwise,
he will apply the decoding operation \cite{NIELSEN_1998,CAVES_1999}

\begin{equation}
R(E(\rho_\E))=\sum_jP_i U^\dagger_j E(\rho_\E) U_j P_i+P_N E(\rho_\E) P_N,
\label{decod}
\end{equation}

\noindent where $P_N=I-\sum_jU_jP_iU^\dagger_j$. Substituting (\ref{TPCP})
in (\ref{decod}) we have that $R(E(\rho_\E))=P_i \rho_\E P_i=\rho_\E$,
if $\rho_\E$ belongs to $\C$, recovering in this way any valid message
sent by Alice.  Next Bob has to verify tag appended to the message.
If it belongs to $\V$, he will accept the message, otherwise he will
discard it suspecting a manipulation in the channel.

In the following subsections we shall analyse the security of this scheme
to several attacks. In the forgery attack we shall regard Eve capable
of intercepting the state travelling from Alice to Bob, discard it,
and forge a new tagged message. In the measurement attack, Eve measures
the message in the channel trying to discover the key used.  Finally,
in the unitary attack, we shall assume that Eve can modify the state in
the channel by means of a unitary operation.  For simplicity we shall
initially consider a two-dimensional tag space $\T$.

\subsection{Forgery attack}

Suppose Eve prepares a forged tagged message $\rho^E_\E$ and sends it
to Bob trying to impersonate Alice.  When Bob receives this message,
depending on the value of the key, he will either apply to the message
the decoding operation (\ref{decod}) or he will do nothing.  After this
decoding, Bob will reject the message if the measurement on the tag space
$\T$ shows that it belongs to $\VP$. Therefore, the probability $P_f$
that Eve deceives Bob is

\begin{equation}
P_f= {1\over 2} {\rm tr}_\E[P_i \rho^E_\E+P_i R(\rho^E_\E)].
\label{probfalso}
\end{equation}

\noindent This quantity depends both on Eve's strategy and on the reverse
of the TPCP map actually chosen by Alice and Bob. She will succeed with
probability one if

\begin{equation}
\rho^E_\E\in\C, 
\label{primera}
\end{equation}

\noindent and

\begin{equation}
R(\rho^E_\E)=\sum_jP_i U^\dagger_j \rho^E_\E U_j P_i + P_N \rho^E_\E
P_{N}\in\C.
\label{segunda}
\end{equation}

To better study these conditions we will make use of the following
decomposition. Defining the orthogonal projection operators $P_i$ and
$P_o$ as the ones that, respectively, project a state from $\E$ onto $\C$
or onto $\CP$, an arbitrary operator $A_\E$ can be written as

\begin{equation}
A_\E= A_{ii}+ A_{io}+ A_{oi}+ A_{oo},
\label{DESCOMPOSICION_IO}
\end{equation}

\noindent where $A_{jk}=P_jA_\E P_k$, with $j,k=i,o$. If the decomposition
(\ref{DESCOMPOSICION_IO}) is used in operator expressions of the form
$\chi_\E=A_\E \rho_\E A_\E^\dagger$, the corresponding `i-o' operators
are related by the matrix equation

\begin{equation}
\left(
\begin{array}{cc}
\chi_{ii} & \chi_{io}\\
\chi_{oi} & \chi_{oo}\\
\end{array}
\right)
=
\left(
\begin{array}{cc}
A_{ii} & A_{io}\\
A_{oi} & A_{oo}\\
\end{array}
\right)
\left(
\begin{array}{cc}
\rho_{ii} & \rho_{io}\\
\rho_{oi} & \rho_{oo}\\
\end{array}
\right)
\left(
\begin{array}{cc}
A_{ii}^\dagger & A_{oi}^\dagger\\
A_{io}^\dagger & A_{oo}^\dagger\\
\end{array}
\right),
\end{equation}

\noindent where $A_{jk}^\dagger=P_k A_\E^\dagger P_j$.

Using this decomposition, condition (\ref{primera}) requires the
following matrix form for $\rho_\E^E$:

\begin{equation}
\rho^E_\E=
\left(
\begin{array}{cc}
\rho^E_\S & 0\\
0 & 0\\
\end{array}
\right).
\label{falso}
\end{equation}

\noindent In condition (\ref{segunda}), the summation in the right
side is always in $\C$ due to the action of $P_i$. We are left with
the second term.  Writing $U_j=U^j_{ii}+U^j_{io}+U^j_{oi}+U^j_{oo}$,
where $U^j_{kl}=P_k U_j P_l$, we have:

\begin{equation}
P_N=I-\sum_jU_j P_i U^\dagger_j=
\left(
\begin{array}{cc}
I & 0\\
0 & I\\
\end{array}
\right)-\sum_j
\left(
\begin{array}{cc}
U^j_{ii} U^{j\dagger}_{ii} & U^j_{ii} U^{j\dagger}_{oi}\\
U^j_{oi} U^{j\dagger}_{ii} & U^j_{oi} U^{j\dagger}_{oi}\\
\end{array}
\right)\equiv
\left(
\begin{array}{cc}
I & 0\\
0 & I\\
\end{array}
\right)-
\left(
\begin{array}{cc}
G_{ii} & H\\
H^\dagger & G_{oi}\\
\end{array}
\right).
\label{igual}
\end{equation}
       
\noindent With this notation the second term of (\ref{segunda})
vanishes when

\begin{equation}
H^\dagger \rho^E_\S H=0, 
\label{mat1}
\end{equation}

\noindent and

\begin{equation}
(G_{ii}-I) \rho^E_\S H=0, 
\label{mat2}
\end{equation}

\noindent where $H=\sum_jU^j_{ii} U^{j\dagger}_{oi}$, and
$G_{ii}=\sum_jU^j_{ii} U^{j\dagger}_{ii}$.  In order to further analyse
these equations let us first study the structure of the $U_j$ operators.
Since the tag space is two-dimensional, each $U_j$ will be a $4\times
4$ square matrix. Given their unitarity, their rows and columns will be
orthonormal vectors. We may write two of these operators in column-vector
form as

\begin{eqnarray}
U_1=(u_1| u_2| u_3| u_4),\\
U_2=(v_1| v_2| v_3| v_4),
\end{eqnarray}

\noindent where the sets $\{u_j\}$ and $\{v_j\}$ are two basis of
a four-dimensional vectorial space. The reversibility condition,
Eq. (\ref{cond2}), imposes that the first two vectors of each set must
also be orthogonal, so the set $\{u_1,u_2,v_1,v_2\}$ will also be a
base. This means that we can only have two operators composing the TPCP
map. A third one will necessarily have its first two columns orthogonal
to the first two columns of the other two operators, which is impossible.
Therefore, our encoding operation can only be composed of up to ${\rm
dim}(\E)/{\rm dim}(\C)$ (2 in our case) unitary operators.

With two unitary operators forming the encoding operation and
a two-dimensional tag space, we can calculate $H=\sum_jU^j_{ii}
U^{j\dagger}_{oi}$. Using the fact that $\{u_1,u_2,v_1,v_2\}$ is an
orthonormal base, we obtain $H=0$, and analogously $G_{ii}=\sum_jU^j_{ii}
U^{j\dagger}_{ii}=I$. So Eve will always fulfil conditions (\ref{mat1})
and (\ref{mat2}). This will always happen if we choose the maximum
number of unitary operators allowed by the dimensions of $\E$ and $\C$.
One possibility to overcome this would be to choose only one unitary
operator, but then the encoding operation would be unitary, and it has
been shown in \cite{CURTY_2002} that with a one-bit key a unitary encoding
is not secure. Another possibility would be to use a larger tag space $\T$
that will increment ${\rm dim}(\E)$ without increasing ${\rm dim}(\C)$
($\V$ will remain as a one-dimensional subspace of $\T$). This will
allow us to use more unitary operators to form the non-unitary TPCP map
without choosing the maximum of operators allowed by the dimensions of
$\E$ and $\C$. Specifically, if we use two qubits for the tag, we have a
four-dimensional tag space $\T$. Imposing that the valid tag subspace $\V$
is one-dimensional, we have that ${\rm dim}(\E)=8$, ${\rm dim}(\C)=2$,
and ${\rm dim}(\E)/{\rm dim}(\C) =4$.  We can thus choose up to three
unitary operators to construct the TPCP map.  With two or three operators
it is easy for Alice and Bob to make impossible for Eve to satisfy
(\ref{mat1}) and (\ref{mat2}).  This follows from the fact that to
fulfil (\ref{mat1}) Eve has to choose $\rho^E_\S$ with support on the
kernel of $H^\dagger$. But since since ${\rm dim}(\C)<{\rm dim}(\CP)$
and $H^\dagger$ acts from $\C$ on $\CP$, Alice and Bob can always choose
$U_j$ such that the kernel of $H^\dagger$ is empty. Therefore, there is
no $\rho_\S^E$ fulfilling (\ref{mat1}). Therefore, in the following we
will shall consider a tag space of dimension greater than two.

\subsection{Measurement attack}

Now Eve tries to measure the message sent by Alice to discover the key
used and replace the message with one of her own. Since Eve does not know
which message, $\rho_\S$, was sent, the only way she has to discern between
the two possible values of the key is measuring the tag. Therefore, to
make this strategy successful, Eve needs the tag of any message encoded
with $E(\cdot)$ to belong to $\VP$ so as to distinguish it from the tag
of a message encoded with $I$ that belongs to $\V$. Specifically, Eve
needs that

\begin{equation}
\sum_j d_j
\left(
\begin{array}{cc}
U^j_{ii} & U^j_{io}\\
U^j_{oi} & U^j_{oo}\\
\end{array}
\right)
\left(
\begin{array}{cc}
\rho_\S & 0\\
0 & 0\\
\end{array}
\right)
\left(
\begin{array}{cc}
U^{j\dagger}_{ii} & U^{j\dagger}_{oi}\\
U^{j\dagger}_{io} & U^{j\dagger}_{oo}\\
\end{array}
\right)
=
\left(
\begin{array}{cc}
0 & 0\\
0 & \rho'_\S\\
\end{array}
\right).
\end{equation}

\noindent This condition requires that

\begin{equation}
\sum_j d_j U^j_{ii} \rho_\S U^{j\dagger}_{ii}=0, \quad \forall \rho_\S,
\label{m1}
\end{equation}

\noindent and 

\begin{equation}
\sum_j d_j U^j_{oi} \rho_\S U^{j\dagger}_{ii}=0, \quad \forall \rho_\S.
\label{m2}
\end{equation}

\noindent Since the $d_j$ are positive, each term in (\ref{m1}) is
positive and Eve needs that

\begin{equation}
U^j_{ii} \rho_\S U^{j\dagger}_{ii}'=0,
\quad \forall \rho_\S, \forall j.
\end{equation}

\noindent This is true $\forall \rho_\S$ iff $U^j_{ii}=0$, $\forall j$,
which is easily avoided by design.

\subsection{Unitary attack}

Let us assume now that Eve performs a unitary quantum operation $F_\E$
on the encoded tagged message in transit between Alice and Bob. This
operation changes the state of the encoded message from $\rho_\E(k)$
to $F_\E \rho_\E(k) F_\E^\dagger$. Bob, ignorant about this action, will
perform his decoding operation on the encoded tagged message received.
Since Eve does not know the value of the key used, the probability 
of deceiving Bob will be

\begin{equation}
P_u= {1\over 2} {\rm tr}_\E[P_i F_\E \rho_\E F_\E^\dagger+
P_i R(F_\E E(\rho_\E) F_\E^\dagger)]
\label{unit},
\end{equation}

\noindent with 

\begin{equation}
R(F_\E E(\rho_\E) F_\E^\dagger)=\sum_k P_i U^\dagger_k F_\E \sum_j d_j U_j \rho
_\E U^\dagger_j F^\dagger_\E U_k P_i + 
P_N F_\E \sum_j d_j U_j \rho_\E U^\dagger_j F^\dagger_\E P_N.
\label{horror}
\end{equation}

\noindent In other words, for Eve to succeed with $P_u=1$, she needs that

\begin{equation}
F_\E \rho_\E F_\E^\dagger\in\C, 
\label{primera2} 
\end{equation}

\noindent and

\begin{equation}
R(F_\E E(\rho_\E) F_\E^\dagger) \in\C. 
\label{segunda2}
\end{equation}

\noindent The first condition means that

\begin{equation}
\left(
\begin{array}{cc}
F_{ii} & F_{io}\\
F_{oi} & F_{oo}\\
\end{array}
\right)
\left(
\begin{array}{cc}
\rho_\S & 0\\
0 & 0\\
\end{array}
\right)
\left(
\begin{array}{cc}
F^{\dagger}_{ii} & F^{\dagger}_{oi}\\
F^{\dagger}_{io} & F^{\dagger}_{oo}\\
\end{array}
\right)
=
\left(
\begin{array}{cc}
\rho'_\S & 0\\
0 & 0\\
\end{array}
\right).
\end{equation}

\noindent This condition is satisfied if

\begin{equation}
F_\E=
\left(
\begin{array}{cc}
F_{ii} & 0\\
0 & F_{oo}\\
\end{array}
\right). \label{fdentro}
\end{equation}

\noindent Writing (\ref{segunda2}) as in (\ref{horror}), we see that the
first term of (\ref{horror}) always fulfils (\ref{segunda2}). Suppose
Eve can choose the operator $F_\E$ such that it commutes with $P_N$,
then the second term in (\ref{horror}) transforms to

\begin{equation}
F_\E P_N \sum_j d_j U_j P_i \rho_\E P_iU^\dagger_j P_N F^\dagger_\E,
\label{feo}
\end{equation}

\noindent where we have also used the fact that $\rho_\E$ belongs to
$\C$, and so $\rho_\E= P_i \rho_\E P_i$. Now using the reversibility
condition (\ref{cond2}) it is easy to see that the operator $P_N U_j P_i$
is identically zero for all $j$:

\begin{equation}
 P_N  U_j P_i = (I - \sum_k U_k P_i U_k^\dagger) U_j P_i = U_j P_i - \sum_{k \neq j} U_k P_i U_k^\dagger U_j P_i - U_j P_i U_j^\dagger U_j P_i=0,
\label{feo3}
\end{equation}

\noindent where the second term is zero because of (\ref{cond2}),
and the other two terms cancel out each other.  Thus we have shown
that, provided Eve can find a unitary operator $F_\E$ fulfilling
(\ref{fdentro}) and commuting with the projector $P_N$, she can deceive
Bob with probability equal to one. Does such unitary operator exist?

Let us explore the similarity of this problem to the one we faced in
\cite{PEREZ_2003} (unitary encodings).  Let $P_M$ be the the projector
orthogonal to $P_N$:

\begin{eqnarray}
P_M= I - P_N = \sum_j U_j P_i U_j^\dagger =
\left(
\begin{array}{cc}
G_{ii} & H\\
H^\dagger & G_{oi}\\
\end{array}
\right). \\
\end{eqnarray}

\noindent Since $F_\E$ is unitary, condition (\ref{fdentro}) is equivalent
to $[F_\E,P_i]=0$, and so our problem (Eve finding $F_\E$ making $P_u=1$)
can be stated in compact form as:

\begin{eqnarray}
&&[F_\E, P_i]=0, \label{recond1} \\
&&[F_\E, P_M]= 0. \label{recond2}
\end{eqnarray}

Note that these equations do not explicitly depend on the index $j$,
that is, on the individual form of the encoding operators $U_j$, but on
a {\it global} characteristic of the TPCP map (through the sum inside
$P_M$). For this individual dependence to dissappear, and also for $P_M$
to be a projector, it is essential the relation between the $U_j$
defined by the reversibility condition.

From \cite{PEREZ_2003} we know that the problem given by
Eqs. (\ref{recond1})--(\ref{recond2}) has solution regardless its
dimension and the explicit form or rank of the two projectors. In fact,
the problem is completely equivalent to the one studied there for the
case of a unitary encoding with just one bit of key. In particular,
the family of $F_\E$ fulfilling (\ref{recond1})-(\ref{recond2}) is
the same, and so Eve can transform the original message as desired.
Thus we conclude that, using the more general set of reversible TPCP
maps for encoding, one bit of key is still not enough to authenticate
one qubit quantum message.

\section{Double encoding}
\label{AUTEN_DOBLE}

In the preceding section we have shown that using more general
encodings, such as TPCP maps, does not provide any advantage over
unitary operations in the authentication of an elemental piece of quantum
information (a qubit). Therefore, it seems reasonable to undertake the
authentication of arbitrary quantum information with unitary encodings.
In \cite{PEREZ_2003} we did so for the case in which Alice and Bob share
a classical $n$-bit key. According to this authentication procedure,
Alice, depending on the value of the key, performs a unitary encoding
rule on the tagged message, $U(k)$, selected from the unitary coding
set $\{U(0),\cdots,U(K-1)\}$, where $K=2^{n}$ and $U(0)=I$. We found the
necessary conditions that this unitary coding set must satisfy to protect
against forgery and unitary attacks.  Unfortunately, these conditions
cannot be fulfilled simultaneously, i.e. the encoding operators that
minimise Eve's probability of success in a forgery attack does not
protect at all against the unitary attack.  In this section we present
a two-stage encoding procedure that tries to circumvent such
deficiency. The different phases of this new protocol are the following:

\begin{enumerate}
\item\textit{Tag 1}: Alice prepares her message, $\rho_{\S}$, which
belongs to the state space $\S$ (${\rm dim}(\S)=S$). She now appends
to the message a tag described by $\rho_{\T_{1}}$ that belongs to the
state space $\T_{1}$ (${\rm dim}(\T_{1})=T_{1}$).  The space of tagged
messages is $\E_{1}=\S\otimes\T_{1}$, and the tagged message will be
described by $\rho_{\E_{1}}=\rho_{\S}\otimes\rho_{\T_{1}}$.  Alice and
Bob openly agree on a particular splitting of the tag space $\T_{1}$
into the direct sum of two subspaces $\T_{1}=\V_{1}\oplus\VP_{1}$,
where $\V_{1}$ is the subspace of valid tags, and $\VP_{1}$ is the
subspace of the invalid ones. This splitting of $\T_{1}$ leads to the
direct sum $\E_{1}=\C_{1}\oplus\CP_{1}$, where $\C_{1}=\S\otimes\V_{1}$
is the subspace of valid messages (${\rm dim}(\C_{1})=C_{1}$), and
$\CP_{1}=\S\otimes\VP_{1}$ is the subspace of the invalid ones (${\rm
dim}(\CP_{1})=D_{1}$).

\item\textit{Encoding 1}: Alice, depending on the value of a
$n_1$-bit key shared with Bob, performs an encoding rule
on the tagged message. The encoding rule, $U_{1}(p)$, is selected
from the unitary encoding set $\{U_{1}(0),\cdots,U_{1}(K_{1}-1)\}$,
where $K_{1}=2^{n_{1}}$ and $U_{1}(0)=I$.  If the tagged message is
$\rho_{\E_{1}}$, the state of the message after this encoding is

\begin{equation}
\rho_{\E_{1}}(p)=U_{1}(p)\rho_{\E_{1}} U^\dagger_{1}(p).
\end{equation}

\noindent Thus, a tagged message encoded by Alice with $U_{1}(p)$
will necessarily belong to $\C_{p}$, the subspace of all the tagged
messages transformed by $U_1(p)$. The aim of this first encoding is to protect against unitary attacks. As it has been shown in
\cite{PEREZ_2003}, the $\C_{k}$ subspaces, $k=0,...,K_1-1$, must be chosen at an
intermediate point between zero and total overlapping so as to offer
the best possible protection against this type of  attack.

\item\textit{Tag 2}: After this first encoding, Alice appends
a second tag $\rho_{\T_{2}}$ to the already encoded and tagged message,
obtaining

\begin{equation}
\rho_{\E}=\rho_{\E_{1}}(p)\otimes\rho_{\T_{2}}.
\end{equation}

\noindent The state space of the second tag is $\T_{2}$ and is
analogously divided into the subspaces of valid and invalid tags
$\T_{2}=\V_{2}\oplus\VP_{2}$. The space of the messages tagged twice is
$\E=\E_{1}\otimes\T_{2}=\S\otimes\T_{1}\otimes\T_{2}$, and the valid
and invalid message subspaces $\E=\C\oplus\CP$ (${\rm dim}(\C)=C$,
${\rm dim}(\CP)=D$). The second tag appended by Alice will thus belong
to $\V_{2}$. Regardless of the dimension of $\T_{2}$, we shall assume
$\V_{2}$ to be one-dimensional, with  $|0\rangle_{\T_{2}}$ a base of
this subspace, and thus $\rho_{\T_{2}}= |0\rangle\langle 0|_{\T_2}$.

\item\textit{Encoding 2}: Finally, Alice performs a second encoding
on the twice-tagged message depending on the value of a second key
$q$ shared with Bob. The encoding rule, $U_{2}(q)$, is selected
from the unitary coding set $\{U_{2}(0),\cdots,U_{2}(K_{2}-1)\}$,
with $U_2(0)=I$. This second encoding is intended to protect the
message against forgery attacks without interfering with the first
encoding.  In order to achieve this goal, Alice and Bob choose a
number of operators $K_{2}-1$ equal to the dimension of the invalid tag
subspace $\VP_2$. Let us consider the following base of this subspace:
$\{|1\rangle_{\T_{2}},|2\rangle_{\T_{2}},\ldots,|K_{2}-1\rangle
_{\T_{2}}\}$.  The encoding operators
$\{U_{2}(0),\cdots,U_{2}(K_{2}-1)\}$ acting on $\E$ can be written as
$U_{2}(k)=\sum_{j,l=0}^{K_{2}-1}U_{2}(k)_{jl}\otimes|j\rangle\langle
l|_{\T_{2}}$ with $U_{2}(k)_{jl}$ acting on $\E_{1}=\S\otimes\T_{1}$. We
shall further impose on the unitary encoding operators the condition
$U_{2}(k)_{0j}=U_{2}(k)_{j0}=\delta_{jk}$. As we shall explain later,
this allows to protect the message against forgery without interfering
with the first encoding. After this second encoding, the expression for
the message to be sent through the channel is:

\begin{equation}
\rho_{\E}(q)=U_{2}(q)  \rho_{\E_{1}}(p)  
\otimes |0\rangle \langle 0|_{\T_2} U^\dagger_{2}(q).
\end{equation}

\item\textit{Verification}: When Bob receives the message, he performs the
matching decoding rule to Alice's second encoding, $U^{\dagger}_{2}(q)$,
and measures the second tag attached. If it belongs to $\V_{2}$,
he will continue, after tracing out the second tag attached, with the
verification procedure: He will first decode, with $U^{\dagger}_{1}(p)$,
Alice's first encoding, and then will measure the first tag and check
whether it belongs to $\V_{1}$.  If this is the case, Bob will regard
the message as authentic and will recover the original plain-text after
tracing out both tags.

\end{enumerate}

We shall now analyse the effects of Eve's attacks on this protocol.

\subsection{Forgery attack}

In this attack Eve has the power to replace the tagged message in
transit between Alice and Bob with a forged tagged message of her
own, $\rho_E$. From this message, Bob will perform his first decoding
obtaining

\begin{equation}
U_{2}^\dagger(q) \rho_E U_{2}(q),
\end{equation}

\noindent with $q$ the value of the second key shared with Alice. The
probability of Eve being undetected is

\begin{equation}
P_f= {1 \over K_2} \sum_{k=0}^{K_{2}-1} \mbox{tr}_{\E} 
\left[P_2 U_2(k)^\dagger
\rho_E U_2(k) \right] =  {1 \over K_2} \sum_{k=0}^{K_{2}-1} 
\mbox{tr}_{\E}
\left[P_2(k) \rho_E \right],
\end{equation}

\noindent where $P_{2}=I_{\E_{1}}\otimes|0\rangle\langle0|_{\T_{2}}$
is the projector onto $\C$, and $P_2(k)=U_2(k)P_2U_2^\dagger(k)$
is the projector onto $\C$ transformed by $U_2(k)$. But, since we
have imposed before that $U_{2}(k)_{0j}=U_{2}(k)_{j0}=\delta_{jk}$,
$P_{2}(k)=I_{\E_{1}}\otimes|k\rangle\langle k|_{\T_{2}}$, and then
$P_f=1/K_2$. Therefore, Alice and Bob can decrease Eve's success
probability increasing $K_2$, protecting in this way the message
against forgery. Note also that the projectors $P_2(k)$ are mutually
orthogonal. In other words, the effect of the condition imposed over the
set of operators $U_2(k)$ is to transform the space of valid messages
$\C$ into  orthogonal, disjoint subspaces. In \cite{PEREZ_2003} we
showed that this type of unitary set provides optimal protection against
forgery attacks.

\subsection{Unitary attack}

Let us now assume that Eve performs a unitary quantum operation $F_\E$
on the encoded tagged message in transit between Alice and Bob. This
operation changes the state of the message from $\rho_{\E}(q)$ to
$F_\E \rho_{\E}(q) F_\E^\dagger$.  Bob, ignorant about this action,
will perform his first decoding operation on the message received,
obtaining as the decoded tagged message

\begin{equation}
U_2^\dagger(q)F_\E \rho_{\E}(q) F_\E^\dagger 
U_2(q).
\end{equation}

\noindent Thus, the probability of Eve being unnoticed after this first
test is

\begin{equation} 
P_u= {1\over K_2} \sum_{k=0}^{K_{2}-1} \mbox{tr}_{\E}
\left[ P_2 U_2^\dagger(k)F_\E \rho_{\E}(k) F_\E^\dagger U_2(k) \right]
= {1\over K_2} \sum_{k=0}^{K_{2}-1} \mbox{tr}_{\E} \left[ F_\E^\dagger  P_2(k) F_\E
\rho_{\E}(k) \right].
\end{equation}
  
\noindent Since the subspaces obtained after transforming $\C$
with the set $U_2(k)$ are all orthogonal, disjoint subspaces of $\E$,
if Eve chooses a unitary operator $F_\E$ acting inside each subspace
separately according to

\begin{equation}
F_\E=\sum_{j=0}^{K_{2}-1}F_{jj}\otimes|j\rangle\langle j|_{\T_{2}},
\end{equation}

\noindent with $F_{jj}$ unitary and acting on $\E_1$, then  $ F_\E^\dagger
P_2(k) F_\E = P_2(k)$, and she will always pass Bob's first test. The
manipulated message that Bob accepts after his first test will be

\begin{equation}
\rho_{\E_1}^\prime (p) = \mbox{tr}_{\T_2} \left( P_2(q) F_\E U_2(q)
\rho_{\E_1}(p) \otimes |0\rangle\langle 0|_{\T_2}U_2^\dagger (q) F_\E^\dagger
\right).
\label{transform1}
\end{equation}

\noindent After some algebra it can be shown that $\rho_{\E_1}^\prime(p)$
reduces to $F_{qq}\rho_{\E_{1}}(p)F^{\dagger}_{qq}$, where
$\rho_{\E_{1}}(p)$ is the original message tagged for the first
time by Alice, and encoded with $U_{1}(p)$. Bob will now decode this
message with $U^{\dagger}_{1}(p)$, and verify the first tag appended
by Alice. Therefore, if Alice and Bob choose the unitary encoding
set $\{U_{1}(0),\cdots,U_{1}(K_{1}-1)\}$ to offer the best possible
protection against a unitary attack regardless of its effectiveness
against a forgery attack, the entire procedure will protect the message
simultaneously against both attacks with the best protection that a
single encoding can give against each attack separately.

\section{Conclusion}
\label{FINALE}

In Section \ref{AUTEN_TPCP} our goal has been to investigate whether the
use of TPCP maps in the encoding and decoding operations could improve the
security of authentication protocols as compared to the unitary encoding
studied in \cite{CURTY_2002} and \cite{PEREZ_2003}. In \cite{CURTY_2002}
we showed that one bit of key is enough to protect one qubit of message
against forgery or measurement attacks, but not against unitary attacks.
One could expect that the extended complexity of a TPCP encoding would
protect more efficiently against these attacks. However, we have found
exactly the opposite to hold: the use of a TPCP map does not improve at
all the security of the protocol with one bit of key; in fact, there
is no reason for Alice and Bob to use it, as it needs more resources.
This result is strongly related to the reversibility condition imposed
on the TPCP map. That condition requires the TPCP map to act inside the
coding subspace, effectively, as a collective unitary operator.

With this idea in mind, in Section \ref{AUTEN_DOBLE} we have approached
the authentication of arbitrary quantum information with a double
encoding. We have shown that it offers simultaneous protection against
unitary and forgery attacks with the best protection that a single
encoding can give to each attack separately.

Many open questions related to the quantum authentication schemes analysed
deserve further investigation. For example, a more practical definition
of the probability of failure is required. For instance, Eve might,
with high probability, transform the original message, but only in a way
such that the fidelity between the original and the transformed state
be still very high. It could also be the case that she could strongly
transform, without being noticed, messages inside a particular subspace
of the valid message space, and be noticed if she transforms messages
outside of it. All these situations would have to be considered by Alice
and Bob in any practical implementation of the protocol.

\section*{Acknowledgements}

The authors acknowledge stimulating discussions with M. Curty.
This work was partially supported by the Spanish Government (grant Nos.\
TIC1999-0645-C05-03, BFM2000-0806 and TIC2001-3217), by Comunidad de
Madrid (Spain, grant No.\ 07T/0063/2000) and Xunta de Galicia (Spain,
grant No.\ PGIDT00PXI322060PR).

\end{document}